\documentclass[aps,showpacs,floatfix,twocolumn,byrevtex,superscriptaddress]{revtex4-1}
\usepackage{amsmath}
\usepackage{amssymb}
\usepackage{amstext}
\usepackage{amsopn}
\usepackage{amsfonts}
\usepackage{amsxtra}
\usepackage[multiple]{footmisc}
\usepackage{bookmark}
\usepackage[english]{babel}
\usepackage{amsmath}
\usepackage{graphicx}
\usepackage{float}
\usepackage{bm}
\usepackage{multirow}
\usepackage{dcolumn}
\usepackage{hyperref}
\usepackage{CJK}
\begin{document}

%%%%%%%%%%%%%%%%%%%%%%%%%%%%%%%%%%%%%%%%%%%%%%%%%%%%%%%%%%%%%%%%%%%%%
%%
%%  Title
%%

\title{Surface skyrmions and dual topological Hall effect in antiferromagnetic topological insulator EuCd$_2$As$_2$}

\author{Min Wu}
\thanks{These authors contributed equally to this work.}
\affiliation{Anhui Province Key Laboratory of Condensed Matter Physics at Extreme Conditions, High Magnetic Field Laboratory,  Chinese Academy of Science, Hefei 230031, China.}
\author{R. Yang}
\email{ryang@seu.edu.cn}
\affiliation{Key Laboratory of Quantum Materials and Devices of Ministry of Education, School of Physics, Southeast University, Nanjing 211189, China}
\author{Xiangde Zhu}
\thanks{These authors contributed equally to this work.}
\affiliation{Anhui Province Key Laboratory of Condensed Matter Physics at Extreme Conditions, High Magnetic Field Laboratory, Chinese Academy of Science, Hefei 230031, China.}
\author{Yixiong Ren}
\thanks{These authors contributed equally to this work.}
\affiliation{Anhui Province Key Laboratory of Condensed Matter Physics at Extreme Conditions, High Magnetic Field Laboratory, Chinese Academy of Science, Hefei 230031, China.}
\affiliation{Department of physics, University of Science and Technology of China. Hefei 230026, China.}
\author{Ang Qian}
\author{Yongjie Xie}
\affiliation{Institute of Physics, Chinese Academy of Sciences, Beijing 100190, China}
\affiliation{School of Physical Sciences, University of Chinese Academy of Sciences, Beijing 100049, China}
\author{Changming Yue}
\affiliation{Department of Physics, Southern University of Science and Technology, Shenzhen 518055, China}
\author{Yong Nie}
\affiliation{Anhui Province Key Laboratory of Condensed Matter Physics at Extreme Conditions, High Magnetic Field Laboratory, Chinese Academy of Science, Hefei 230031, China.}
\author{Xiang Yuan}
\affiliation{Department of Physics, School of Physics and Electronic Science, East China Normal University, Shanghai 200241, China}
\author{Ning Wang}
\affiliation{Anhui Province Key Laboratory of Condensed Matter Physics at Extreme Conditions, High Magnetic Field Laboratory, Chinese Academy of Science, Hefei 230031, China.}
\author{Daifeng Tu}
\author{Ding Li}
\affiliation{Anhui Province Key Laboratory of Condensed Matter Physics at Extreme Conditions, High Magnetic Field Laboratory, Chinese Academy of Science, Hefei 230031, China.}
\affiliation{Department of physics, University of Science and Technology of China. Hefei 230026, China.}
\author{Yuyan Han}
\author{Zhaosheng Wang}
\affiliation{Anhui Province Key Laboratory of Condensed Matter Physics at Extreme Conditions, High Magnetic Field Laboratory, Chinese Academy of Science, Hefei 230031, China.}
\author{Yaomin Dai}
\affiliation{National Laboratory of Solid State Microstructures and Department of Physics, Nanjing University, Nanjing 210093, China}
\affiliation{Collaborative Innovation Center of Advanced Microstructures, Nanjing University, Nanjing 210093, China.}
\author{Guolin Zheng}
\affiliation{Anhui Province Key Laboratory of Condensed Matter Physics at Extreme Conditions, High Magnetic Field Laboratory, Chinese Academy of Science, Hefei 230031, China.}
\author{Jianhui Zhou}
\email{jhzhou@hmfl.ac.cn}
\affiliation{Anhui Province Key Laboratory of Condensed Matter Physics at Extreme Conditions, High Magnetic Field Laboratory, Chinese Academy of Science, Hefei 230031, China.}
\author{Wei Ning}
\email{ningwei@hmfl.ac.cn}
\affiliation{Anhui Province Key Laboratory of Condensed Matter Physics at Extreme Conditions, High Magnetic Field Laboratory, Chinese Academy of Science, Hefei 230031, China.}
\author{Xianggang Qiu}
\affiliation{Institute of Physics, Chinese Academy of Sciences, Beijing 100190, China}
\affiliation{School of Physical Sciences, University of Chinese Academy of Sciences, Beijing 100049, China}
%\affiliation{Songshan Lake Materials Laboratory, Dongguan, Guangdong 523808, China}
\author{Mingliang Tian}
\affiliation{Anhui Province Key Laboratory of Condensed Matter Physics at Extreme Conditions, High Magnetic Field Laboratory, Chinese Academy of Science, Hefei 230031, China.}
\affiliation{Collaborative Innovation Center of Advanced Microstructures, Nanjing University, Nanjing 210093, China.}
\affiliation{Department of Physics, School of Physics and Materials Science, Anhui University, Hefei 230601, Anhui, China.}
\date{\today}

%%%%%%%%%%%%%%%%%%%%%%%%%%%%%%%%%%%%%
%%
%% Abstract
%%
%

\begin{abstract}
In this work, we synthesized single crystal of EuCd$_2$As$_2$, which exhibits A-type antiferromagnetic (AFM) order with in-plane spin orientation below $T_N$ = 9.5~K.
Optical spectroscopy and transport measurements suggest its topological insulator (TI) nature with an insulating gap around 0.1~eV.
Remarkably, a dual topological Hall resistivity that exhibits same magnitude but opposite signs in the positive to negative and negative to positive magnetic field hysteresis branches emerges below 20~K.
With magnetic force microscopy (MFM) images and numerical simulations, we attribute the dual topological Hall effect to the N\'{e}el-type skyrmions stabilized by the interactions between topological surface states and magnetism, and the sign reversal in different hysteresis branches indicates potential coexistence of skyrmions and antiskyrmions.
Our work uncovers a unique two-dimensional (2D)  magnetism on the surface of intrinsic AFM TI, providing a promising platform for novel topological quantum states and AFM spintronic applications.
\end{abstract}

%  72.15.-v  Electronic conduction in metals and alloys
%  74.70.-b  SC: Superconducting materials other than cuprates
%  78.20.-e  Optical properties of bulk materials and thin films
%  78.30.-j  Infrared and Raman spectra

%\pacs{72.15.-v, 74.70.-b, 78.30.-j}
\maketitle
%
%%%%%%%%%%%%%%%%%%%%%%%%%%%%%%%%%%%%%%%%%%%%%%%%%%%%%%%%%%%%%%%%%%%%%%%%%%%%%%%
\section{Introduction}
%%%%%%%%%%%%%%%%%%%%%%%%%%%%%%%%%%%%%%%%%%%%%%%%%%%%%%%%%%%%%%%%%%%%%%%%%%%%%%%
The entanglement between topology and magnetism can generate various novel quantum phenomena such as quantum anomalous Hall effect, axion insulator states, and skyrmions, among which the skyrmions, for their small size, low energy consumption, and high mobility with low current densities, are regarded as promising candidates for next-generation data storage and computing devices~\cite{Fert2017, He2022}. Realizing the skyrmions and understanding the underlying mechanism in various materials is vital in condensed matter physics and applications~\cite{Fert2017, Wu2020, Jiang2020NM, Yasuda2016}.
Recently, the skyrmions have been discovered on the interface of TI/magnet heterostructures and were attributed to the asymmetric Dzyaloshinsky-Moriya (DM) interactions between magnetic moments delivered by electron topological state\cite{He2022, Yasuda2016}.
Nevertheless, the precisely controlled thickness, lattice matching, and homogeneity of dopant concentration poses significant challenges for the large-scale application of skyrmions in 2D systems.
Three-dimensional (3D) magnetic TIs harbor intrinsic magnetic order, insulating bulk states and the massless Dirac surface states that are characterized by large spin-momentum locking and the nontrivial Berry phase~\cite{Hasan2010RMP,Qi2011RMP, Hsieh2008Nature, He2011,LiuMH2012, Tokura2019NRP,Bernevig2022Nature}.
Whether the intrinsic magnetic TIs can interweave topology and magnetism on the surface, providing us a naturally formed 2D platform for further advancement in topological quantum phenomena, including skyrmions, is still an open question.

Recently, the van der Waals magnets [MnBi$_2$Te$_4$][Bi$_2$Te$_3$]$_n$ family were discovered as an intrinsic AFM TI~\cite{Otrokov2019,YangSQ2021PRX,Ovchinnikov2021NL,ZangZH2022PRL},
exhibiting exotic topological phases of matter, such as the quantum anomalous Hall effect~\cite{DengYJ2020science}, axion state~\cite{Liu2020NM} and electrically controlled layered Hall effect~\cite{Gao2021Nature, Gao2023, Wang2023}.
Nevertheless, these phenomena are primarily observed in thin flakes and are sensitive to the thickness, limiting their widespread implementation.
Furthermore, MnBi$_2$Te$_4$ is metallic in bulk, the entanglement between bulk and surface states at the Fermi level hinders the study of interactions between topology and magnetism.

EuCd$_2$As$_2$ has been proposed as another candidate for AFM TI~\cite{Rahn2018, Soh2019}. It crystallizes in a layered crystal structure with space group P-3$m$1 (No. 164).
The Cd$_2$As$_2$ bilayer and triangular Eu layers are mutually staggered in the unit cell and contribute to the low-energy bands and local moments ($4f$ electrons), respectively~\cite{Ma2019}.
Below $T_N =$ 9.5~K, Eu's magnetic moments form an A-type AFM order (in-plane FM coupling and interlayer AFM coupling, inset of Fig.~\ref{fig:magnetoresistivity}(a)).
For the weak magnetic anisotropic energy, the magnetic ground state with moments oriented to either the \emph{c}-axis or the \emph{ab}-plane becomes possible~\cite{Soh2019, Gati2021}.
With the magnetic moments along the \emph{c}-axis, Dirac semimetal state was observed by the angle-resolved photoelectron spectroscopy studies~\cite{Ma2019, Ma2020}.
When the magnetic moments lie in the \emph{ab}-plane~\cite{Soh2019}, for the \emph{PT} symmetry~\cite{Otrokov2019}, EuCd$_2$As$_2$ was predicted as an AFM TI~\cite{Soh2019, He2022, Wang2019}.
This unique system will provide a unique platform to study the interplay between topological surface state and antiferromagnetism without the affections from the bulk bands~\cite{Soh2019, Jo2020}.
In this work, we synthesized the single crystal of insulating EuCd$_2$As$_2$ and conducted a comprehensive investigation of its properties through transport, magnetism, optical measurements, and theoretical calculations(see the methods in section I of the supplementary materials (SM)~\cite{Supplementary}).

%%%%%%%%%%%%%%%%%%%%%%%%%%%%%%%%%%%%%%%%%%%%%%%%%%%%%%%%%%%%%%%%%%%%%%%%%%%%%%%%%%%
%Results
%%%%%%%%%%%%%%%%%%%%%%%%%%%%%%%%%%%%%%%%%%%%%%%%%%%%%%%%%%%%%%%%%%%%%%%%%%%%%%%

\section{Results}

%%%%%%%%%%%%%%%%%%%%%%%%%%%%%%%%%%%%%%%%%%%%%%%%%%%%%%%%%%%%%%%%%%%%%%%%%%%%%%%%%%%
%Magneto resistivity
%%%%%%%%%%%%%%%%%%%%%%%%%%%%%%%%%%%%%%%%%%%%%%%%%%%%%%%%%%%%%%%%%%%%%%%%%%%%%%%
% Figure 2
%
\begin{figure}[tb]
\centerline{
\includegraphics[width=1\columnwidth]{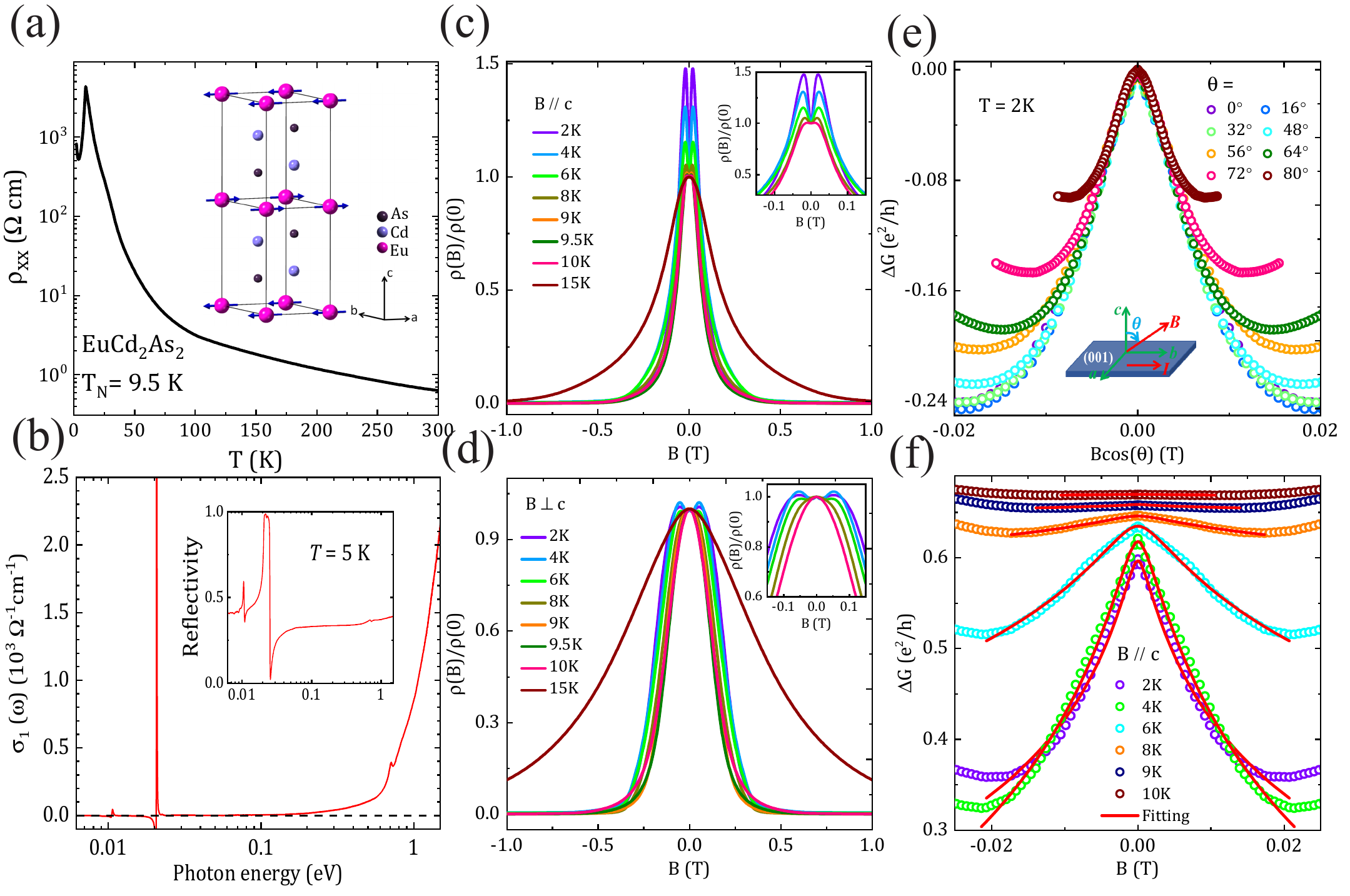}
}
\caption{%\textbf{Negative magnetoresistivity and weak antilocalization effect.}
(a) $T$-dependent resistivity at $B=0$~T. Inset is the crystal and magnetic structure. (b) is EuCd$_2$As$_2$'s optical conductivity (absorption) at 5 K, inset is the corresponding reflectivity. (c) and (d) are normalized magnetoresistivity (MRs) measured at various $T$s for $B\parallel c$ and  $B\perp c$. (d) Magnetoconductances (MCs) as a function of the normal component of magnetic field $B$ at different angles ($T=2$~K). Inset: schematic illustration of the experimental geometry. (e) The MC curves at selected $T$ can be well fitted by HLN formula at the low magnetic fields (red lines).}
\label{fig:magnetoresistivity}
\end{figure}

%\textbf{Transport measurements: negative magnetoresistivity (MR) and weak antilocalization effect.}
Firstly, the magnetic measurements reveled its A-type AFM order with in-plane spin configuration below $T_N=$ 9.5~K (inset of Fig.~\ref{fig:magnetoresistivity}a, see the detail of the measurements in section I of SM~\cite{Supplementary}).
To confirm its TI nature, we further carried out the electrical transport measurements.
In Fig.~\ref{fig:magnetoresistivity}a, in contrast to previous studies which revealed typical metallic features~\cite{Ma2019, Ma2020}, our sample exhibits a large resistivity (0.7 $\sim$8~$\Omega$ cm) at room temperature that increases by 4$\sim$5 orders upon cooling to 15~K, reflecting typical semiconducting behavior (Fig.~\ref{fig:magnetoresistivity}a).
This is further corroborated by the optical spectroscopy (Fig.~\ref{fig:magnetoresistivity}b and section II of SM~\cite{Supplementary}), in which the intraband metallic response is absent with an insulating gap around 0.1~eV~\cite{Dressel2002}.
Below $T_N$, the resistivity starts to drop, giving rise to a sharp peak (Figs.~\ref{fig:magnetoresistivity}a).
Under magnetic fields, the resistivities are greatly suppressed for both $B\parallel$c and $B\perp$c configurations (Figs.~\ref{fig:magnetoresistivity}c and d, and Figs. S4 and S5 in SM~\cite{Supplementary}).
In the AFM state, the magnetoresistivities (MRs) are suppressed until 0.5~T, far below the saturation field of magnetization in both directions (Fig. S1d in SM~\cite{Supplementary}).
Above $T_N$, the negative MR becomes moderate but extends to broader field range.
Since considerable FM fluctuations are revealed by the magnetic susceptibility (Fig. S1d in SM~\cite{Supplementary}), one can ascribe the negative MR to the suppression of spin fluctuations.
In the AFM ordered state, the residual fluctuations can be quickly suppressed below 0.5~T, whereas above $T_N$, more significant fluctuations can persist up to higher fields.
Notably, in Fig. S4 of SM~\cite{Supplementary}, with the suppressed FM fluctuations under magnetic field, the resistivity peak indicating the AFM transition shifts to higher $T$s, reflecting the competition between FM and AFM interactions~\cite{Rahn2018}, which will arouse the magnetic instability and skyrmions.

%%%%%%%%%%%%%%%%%%%%%%%%%%%%%%%%%%%%%%%%%%%%%%%%%%%%%%%%%%%%%%%%%%%%%%%%%%%%%%%%%%%
%Topological Hall effect and MFM measurement
%%%%%%%%%%%%%%%%%%%%%%%%%%%%%%%%%%%%%%%%%%%%%%%%%%%%%%%%%%%%%%%%%%%%%%%%%%%%%%%
%Figure 3
%
\begin{figure*}[tb]
\centerline{
\includegraphics[width=1.6\columnwidth]{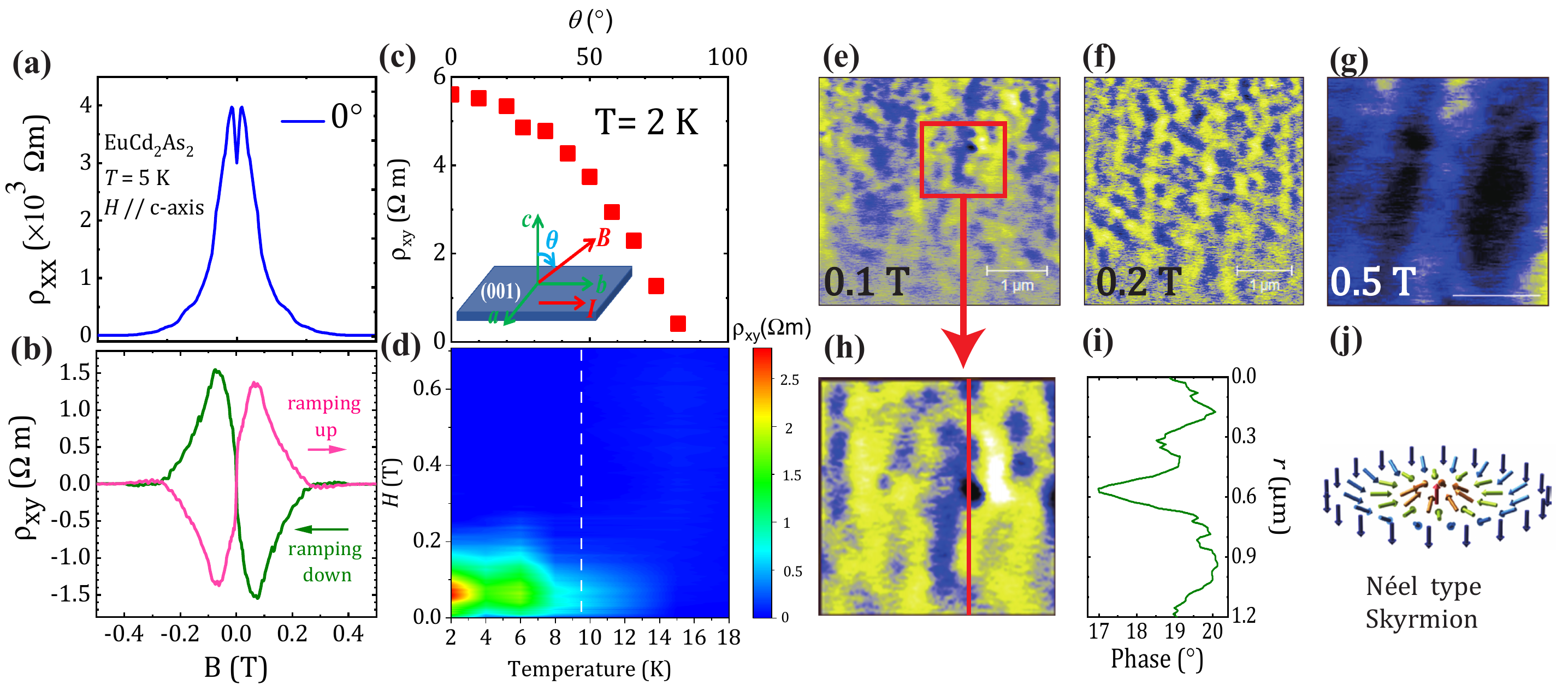}
}
\caption{%\textbf{Nonlinear Hall resistivity and MFM images on the (001) surface of EuCd$_2$As$_2$.}
(a) and (b) are MR and Hall resistivities of EuCd$_2$As$_2$ with the magnetic field along the c-axis. The green and pink arrow represents the positive to negative and negative to positive magnetic field hysteresis branches, respectively. (c) is the angle-dependent magnitude of NHR at 2 K ($\theta$ represents the angle between the magnetic field and the c-axis as shown in the inset). (d) Intensity plot of NHR for H// c at various magnetic fields and temperatures, the white dashed line denotes the $T_N$. (e-g)are the spatial-resolved phase (contrast), which reflects the magnetization (see Fig. S8 of SM~\cite{Supplementary} for more details), measured at 0.1, 0.3, and 0.5~T, more results are shown in Fig. S9 of SM~\cite{Supplementary}). (h) is the enlarged view of the red box in (e). (i) is the phase distribution along the incision (red line) in (h). (j) describes the spin texture of a typical N\`{e}el-type skyrmion.}
\label{fig:THE and MFM}
\end{figure*}

%%%%%%%%%%%%%%%%%%%%%%%%%%%%%%%%%%%%%%%%%%%%%%%%%%%%%%%%%%%%%%%%%%%%%%%%%%%%%%%%%%%%%%%%%%%%%%%%
%WAL&surface state
Besides the negative MR, remarkable positive MRs were also observed for in- and out-of-plane fields below 0.1~T, resulting in a sharp dip centered at 0~T (Figs.~\ref{fig:magnetoresistivity}d and e, Figs. S5 and S6 in SM~\cite{Supplementary}), corresponding to obvious cusp-like structures in the magnetoconductance (MC) below $T_N$ (Figs.~\ref{fig:magnetoresistivity}e and f).
Figure~\ref{fig:magnetoresistivity}d summarizes the curves of MCs ($\Delta G=1/R(B)-1/R(0)$) as a function of projected field along the c-axis $B\cos\theta$ ($\theta$ is the angle between the magnetic field and the c-axis) at $T =$ 2~K.
It is evident that all curves exhibit cusp-like structures and converge to the same tendency as they approach 0~T.
In Fig.~\ref{fig:magnetoresistivity}f, we fit the negative MCs in terms of the Hikami-Larkin-Nagaoka (HLN) formula (red curves)~\cite{He2011, Hikami1980}:~$\Delta \sigma (B)= \sigma (B)-\sigma (0)=\frac{\alpha e^2}{2\pi^2 \hbar}[\psi(\frac{1}{2}+\frac{\hbar}{4eBl^{2}_{\phi}})-ln(\frac{\hbar}{4eBl^{2}_{\phi}})]$, where $l_{\phi}$ is the phase coherent length, $\phi$ is the digamma function, and $\alpha$ is the weak anti-localization (WAL) coefficient, for which topological surface state will give the value around -0.5~\cite{Bao2012}.
At $T$=2~K, the HNL formula provides an excellent fit to the negative MC and yields $\alpha$=-0.45 and $l_{\phi}$=563~nm, suggesting the 2D nature of WAL in our EuCd$_2$As$_2$ sample.
This behavior is commonly observed in TIs and attributed to the WAL effect, which is considered as the hallmark of topological surface states due to strong spin-orbit coupling in bulk~\cite{Chen2011, Zhang2020, LiuMH2012, He2011}.
Given that our sample exhibits insulating behavior in bulk, these results suggest the gapless Dirac surface states residing in the band gap.
Furthermore, by checking the MC curves in Figs.~\ref{fig:magnetoresistivity}e and f, we notice that WAL can be suppressed by increasing either $T$ above $T_N$ or the magnetic field strength, implying a strong correlation between the topological insulating state and the long-range AFM order.

%\textbf{Nonlinear Hall resistivity.}
Since the interplay between electronic topology and magnetism can give rise to chiral spin textures, such as spiral magnetic order and skyrmions, leading to topological Hall effect, to uncover the exotic electromagnetic responses in our insulating EuCd$_2$As$_2$, we further measured the Hall resistivities at low $T$s.
As shown in Fig.~\ref{fig:THE and MFM}b and Fig.~S7 in SM~\cite{Supplementary}, the absent linear Hall resistivity reflects the bulk insulating nature.
Nevertheless, a remarkable nonlinear Hall resistivity (NHR) emerges below 0.5~T at $T<$20~K.
Unexpectedly, the sign of NHR can be inverted by simply reversing the direction of field sweeping~(Fig. S7 in SM~\cite{Supplementary}).
Rotating the external field from the \emph{c}-axis to the \emph{ab}-plane (Fig.~\ref{fig:THE and MFM}c) gradually suppresses the NHR to zero.
Since the out-of-plane field will cant the spin to the $c$-axis, it is evident that the non-coplanar spin configuration is indispensable for generating the NHR~\cite{Yokouchi2014}.
As shown by Fig.~\ref{fig:THE and MFM}d and Fig. S7 in SM~\cite{Supplementary}, the NHR predominantly exists below 20~K, grows rapidly below $T_N$, and attains its maximum at the lowest $T$ (2~K).
Note that similar NHRs were also observed in metallic EuCd$_2$As$_2$ and attributed to the momentum-space Berry curvature associated with the magnetic-field-induced topological phase transition~\cite{Cao2022}.
However, the situation in our sample differs from previous studies in several aspects: i) the NHR develops without bulk itinerant carriers; ii) no insulator-to-metal transition is observed in the magneto-optical measurements up to 8~T~\cite{Caimi2006}(Fig.~S3 of SM~\cite{Supplementary}), at which the moments are fully polarized (Fig.~S1d of SM~\cite{Supplementary}), excluding the possibility of a topological phase transition; iii) the sign reversal of NHR in different field-sweeping directions is inconsistent with the anomalous Hall effect driven by Berry curvature in momentum space.
Nevertheless, the peak-like NHR resembles the topological Hall effect induced by the real-space Berry curvature arising from either spiral spin order or the skyrmions~\cite{Yasuda2016, Fert2017, Nakajima2017, Wu2020, WangL2018, Matsuno2016}.
This is consistent with NHR's perpendicular anisotropy (Fig.~\ref{fig:THE and MFM}c), which suggests a non-coplanar spin configuration.

%%%%%%%%%%%%%%%%%%%%%%%%%%%%%%%%%%%%%%%%%%%%%%%%%%%%%%%%%%%%%%%%%%%%%%%%%%%%
%Skyrmion
%%%%%%%%%%%%%%%%%%%%%%%%%%%%%%%%%%%%%%%%%%%%%%%%%%%%%%%%%%%%%%%%%%%%%%%%%%%%

%%%%%%%%%%%%%%%%%%%%%%%%%%%%%%%%%%%%%%%%%%%%%%%%%%%%%%%%%%%%%%%%%%%%%%%%%%%%%%%
%Figure 4 :Topological Hall effect and MFM measurement
%
\begin{figure*}[tb]
\centerline{
\includegraphics[width=1.2\columnwidth]{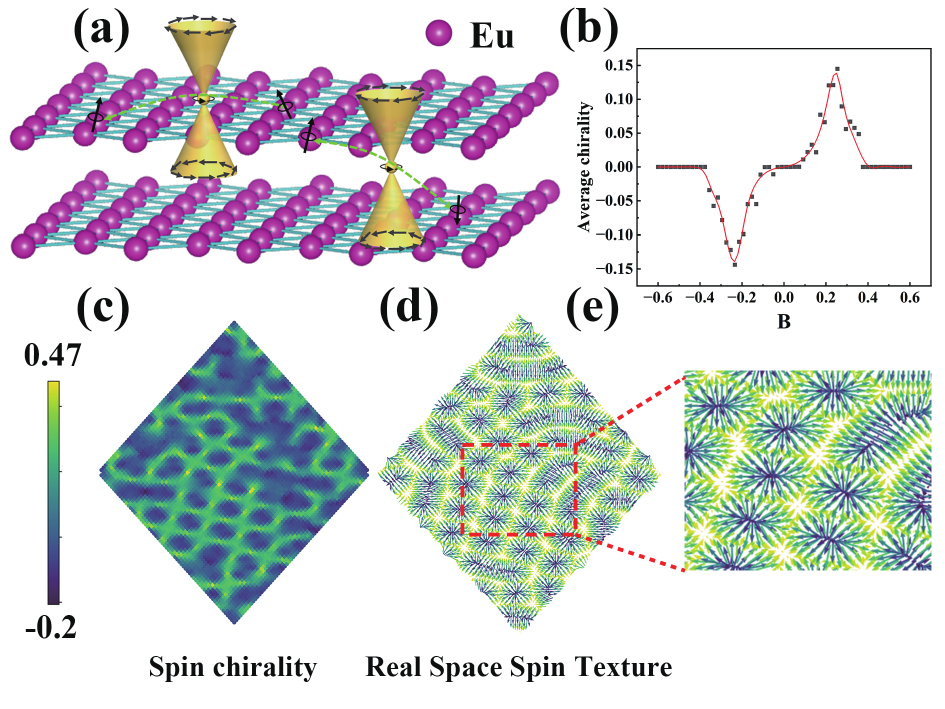}
}
\caption{%\textbf{Skyrmion simulation and the corresponding scalar spin chirality.}
(a) is the schematic diagram of topology-magnetism interaction within the topmost double layer of EuCd$_2$As$_2$. (b) The calculated field-dependent average scalar spin chirality (black square), the red line is the guidance to the eyes. (c) is the spacial distribution of the scalar spin chirality on the topmost layer of EuCd$_2$As$_2$. (d) is the simulated spin texture on the surface of EuCd$_2$As$_2$ based on our topology-magnetism interaction model. (e) is the zoom in of the spin texture in (d), from which one can identify the typical N\'{e}el type skyrmion, which is inline with the observation. }
\label{fig:Theory}
\end{figure*}

%\textbf{Skyrmions observed by MFM image.}
To gain direct evidence of the skyrmions, we measured the spatial-resolved spin texture
through MFM imaging on EuCd$_2$As$_2$'s freshly cleaved (001) surface~\cite{Wu2020, WangL2018, Matsuno2016}.
In Figs.~\ref{fig:THE and MFM} (e-g) and Fig. S9 of SM~\cite{Supplementary}, the measurements were carried out at 5~K ($< T_N$) with a magnetic field along the c-axis ranging from 0 to 0.5~T, where the NHR was observed.
Before the measurements, the sample was cooled down under field 0.5~T.
In MFM image with high spatial resolution, magnetic domains exhibiting opposite magnetization are discerned.
Since our sample is cooling under a magnetic field (0.5~T), this phenomenon is intrinsic.
Unlike conventional labyrinthine magnetic domains~\cite{Rosei2004}, we discovered granular magnetic domains with the size of hundreds of nanometers on EuCd$_2$As$_2$'s surface~(Figs.~\ref{fig:THE and MFM}(e-g)).
As the magnetic field increases up to 0.2~T, the granular domains become more distinct.
However, when the magnetic field exceeds 0.5~T, the spins on the surface are fully polarized with no discernible domain (Fig.~\ref{fig:THE and MFM}g).
We notice that the domain structures exist at field significantly below the saturation field of bulk spins, but vary coherently with the NHR (Fig.~\ref{fig:THE and MFM}b) and negative MR (Fig.~\ref{fig:THE and MFM}a), indicating their intimate relations.
To determine the spin texture within granular magnetic domains, we conducted a detailed examination of a single domain and analyzed its spatial magnetization along the incision depicted in Fig.~\ref{fig:THE and MFM}h.
In Fig.~\ref{fig:THE and MFM}i, the magnetization distribution across this domain displays a symmetric V-shape with opposite polarities at the boundary and center, reminiscent of the superconducting vortex observed on Nb films(Fig.~S8 of SM~\cite{Supplementary}).
Such vortex-like spin configuration is consistent with the N\'{e}el type skyrmion~\cite{Fert2017, Wu2020}.
It is the first observation of the skyrmions on the natural surface of an intrinsic AFM TI.
The magnetization distribution suggests skyrmions with the radius around 0.2~$\mu$m, which is comparable to the ones in TI/magnet heterostructures~\cite{Wu2020}.
Thus, we can ascribe the NHR to the topological Hall effect generated by skyrmions.

%%%%%%%%%%%%%%%%%%%%%%%%%%%%%%%%%%%%%%%%%%%%%%%%%%%%%%%%%%%%%%%%%%%%%%%%%%%%%%%%%%%%%%%%%%%%%%
%Discussion
%%%%%%%%%%%%%%%%%%%%%%%%%%%%%%%%%%%%%%%%%%%%%%%%%%%%%%%%%%%%%%%%%%%%%%%%%%%%%%%%%%%%%%%%%%%%%%%
%
\section{Discussion}

Let us now delve into the origin of skyrmions in AFM TI EuCd$_2$As$_2$.
Previously, skyrmions have predominantly been observed in systems with broken inversion symmetry, which allows for finite DM interactions~\cite{Fert2017}.
However, EuCd$_2$As$_2$'s lattice structure is centrosymmetric.
Although the skyrmions have been observed in centrosymmetric Gd$_2$PdSi$_3$~\cite{Kurumaji2019} and GdRu$_2$Si$_2$~\cite{Yasui2020}, the underlying mechanism that involves Ruderman-Kittel-Kasuya-Yosida (RKKY) and four-spin interactions is mediated by itinerant carriers~\cite{He2018}, while EuCd$_2$As$_2$ is insulating in bulk.
Moreover, the granular domains spanning hundreds of nanometers (Fig.~\ref{fig:THE and MFM}h) are significantly larger than those induced by multiple spin exchange interactions~\cite{Kurumaji2019}.

In metallic EuCd$_2$As$_2$, NHR varies synergistically with negative MR and bulk magnetization, indicating their intimate relations~\cite{Cao2022}.
In our case, as shown in Fig.~\ref{fig:THE and MFM}, NHR and negative MR vary consistently with the surface spin.
Since our EuCd$_2$As$_2$ sample is TI, without the affection from the bulk bands, we speculate that the NHR only exists in the surface layer and originates from the interplay between topological surface states and antiferromagnetism.
Due to the reduced anisotropy energy in thin layers, magnetic moments on the surface can be easily aligned by a much weaker field compared to that required for bulk moments (Fig.~\ref{fig:THE and MFM}g).
This accounts for the significantly reduced field range for both NHR and negative MR compared to the metallic EuCd$_2$As$_2$~\cite{Cao2022}.
Moreover, although EuCd$_2$As$_2$'s lattice is centrosymmetric in bulk, the in-plane spin configuration breaks the $C_3$ symmetry.
Due to strong spin-orbit coupling, the Dirac surface state manifests as a pronounced spin-momentum locking.
Therefore, it is feasible to generate non-coplanar spin textures such as skyrmions due to the DM interactions meditated by the Dirac surface states.

Then, we conduct a numerical simulation based on a toy model with bilayer Eu atoms and Dirac surface state (Fig.~\ref{fig:Theory}a), during which the Dirac surface states mediate either DM interactions between Eu's magnetic moments~\cite{JJZhu2011PRL,HRChang2015PRB} or magnetic interactions as shown in Fig.~\ref{fig:Theory}a (see the detail of the simulation in section VII of the SM~\cite{Supplementary}).
Fig.~\ref{fig:Theory}d displays our simulation of the spin texture on the surface of EuCd$_2$As$_2$.
It verifies that a significant DM interaction in topmost thin layers can indeed lead to N\'{e}el-type skyrmions under moderate out-of-plane magnetic fields.
Further increasing the magnetic field will deform the skyrmions into strip-like chiral domains (Fig. S11 in the SM~\cite{Supplementary}) and finally polarize the whole surface, which is consistent with experimental observations (Figs.~\ref{fig:THE and MFM} e-g).
Correspondingly, in Fig.~\ref{fig:Theory}c, the chiral spin texture contributes finite scalar spin chirality~($\mathbf{S}_{i}\cdot(\mathbf{S}_{j}\times\mathbf{S}_{k})$) in real space, whose ensemble average value is proportional to the topological Hall resistivity~\cite{Zheng2021NC}.
Then, we calculated the average scalar spin chirality in Fig.~\ref{fig:Theory}b and notice that its field dependence is in line with the measured NHR when the field is swiping from negative to positive (pink arrow in Fig.~\ref{fig:THE and MFM}b).
The agreement between simulation and measurements further supports that the NHR observed in insulating EuCd$_2$As$_2$ originates from the topological Hall effect induced by surface skyrmions.

Finally, it is noteworthy that the sign reversal of NHR in increasing and decreasing field procedures (Fig.~\ref{fig:THE and MFM}b) is unique and absent in reported magnetic materials, synthetic heterostructures~\cite{Nagaosa2020NRM, Fert2017}, and even in the metallic EuCd$_2$As$_2$~\cite{Cao2022}, resembling a pair of mirrored topological Hall resistivity, namely the dual topological Hall effect.
Such behavior is similar to the topological Hall effect in Mn$_2$RhSn~\cite{Sivakumar2020}, where magnetic dipolar and anisotropic DM interactions give rise to coexistence of skyrmions and antiskyrmions~\cite{Tong2018}.
Since the topological Hall resistivity is determined by skyrmion's topological charge $N_{sk}=mp$ ($m=\pm 1$ is the vorticity and $p= \pm 1$ represents the polarity~\footnote{$m=$+1 and -1 stand for skyrmion and antiskyrmion, respectively. $p=$ +1 and -1 represent the orientation of the core spin of the skyrmion/antiskyrmion, which is determined by the magnetization of the background.}), at the same field (same magnetization), the opposite signs of NHR in decreasing and increasing field procedures would come from either skyrmions or antiskyrmions, which have opposite vorticities.
On the other side, because of the inconsistency between surface and bulk magnetization under external fields~(Figs.\ref{fig:THE and MFM}g and S1d of SM~\cite{Supplementary}), in EuCd$_2$As$_2$, the coupling between surface and bulk magnetic moments could be the driven mechanism for skyrmions and antiskyrmions~\cite{Tong2018}.
Nevertheless, to pin down the underlying physics, further theoretical and experimental researches are required.

%%%%%%%%%%%%%%%%%%%%%%%%%%%%%%%%%%%%%%%%%%%%%%%%%%%%%%%%%%%%%%%%%%%%%%%%%%%%%%%%%%%%%%%%%%%%%%%
\section{Conclusion}
%%%%%%%%%%%%%%%%%%%%%%%%%%%%%%%%%%%%%%%%%%%%%%%%%%%%%%%%%%%%%%%%%%%%%%%%%%%%%%%%%%%%%%%%%%%%%%%%
In summary, we synthesized the single crystal of EuCd$_2$As$_2$, which shows A-type AFM order with in-plane spin orientation below $T_N= $9.5~K.
With optical spectroscopy and transport measurements, we identified its TI nature with the band gap around 0.1~eV .
%The presence of 2D Dirac surface states was suggested by the WAL effect in the MR, which is consistent with theoretical predictions~\cite{Wang2019}.
Unexpectedly, a unusual dual topological Hall effect develops below 20~K and exhibits different signs in the positive to negative and negative to positive magnetic field hysteresis branches.
Utilizing MFM measurements and theoretical simulations, this anomalous NHR is attributed to the N\'{e}el-type skyrmions induced by the interplay between topological surface states and magnetism, and the sign reversal of NHR indicates the coexistence of skyrmion and antiskyrmion.
Therefore, our findings have unveiled an exotic 2D magnetism that manifests exclusively on the surface of a bulk AFM TI without the affection from the bulk bands.
In contrast to TI/magnet heterostructures, this unique magnetic system provides a new avenue to realize 2D quantum phenomena without complex 2D fabrications, greatly facilitating their application in AFM spintronics.

%
%%%%%%%%%%%%%%%%%%%%%%%%%%%%%%%%%%%%%%%%%%%%%%%%%%%%%%%%%%%%%%%%%%%%%%%%%%%%%%%
%
% Acknowledgment
%
\section{Acknowledgments}
We thank Di Xiao, Matthew Daniels, Congcong Le, Yong Hu, M. Dressel, S. Hayami, S. M. Nie and M. Scheffler for fruitful discussions. This work was supported by the National Key Research and Development Program of China (Grant No. 2021YFA1600201), the Natural Science Foundation of China (No. U19A2093, U2032214, U2032163). R.Yang acknowledge the support from the Alexander von Humboldt foundation.
%
%
%%%%%%%%%%%%%%%%%%%%%%%%%%%%%%%%%%%%%%%%%%%%%%%%%%%%%%%%%%%%%%%%%%%%%%%%%%%%%%%
%
% Author contribution
%
%\section{Author contribution}
%M.W, N.W and X.-D.Z grew the single crystals and carried out the transport measurements. Y.M.D and R.Y. measured the optical spectroscopy. A.Q carried out the MFM measurement. C.-M.Y and Y.X.R performed the first-principles calculations and numerical simulations. R.Y. analyzed the data and prepared the manuscript with comments from all authors. R.Y, J.-H.Z, N.W, and M.-L.T supervised this project.
%
%%%%%%%%%%%%%%%%%%%%%%%%%%%%%%%%%%%%%%%%%%%%%%%%%%%%%%%%%%%%%%%%%%%%%%%%%%%%%%%%
%%
%% Competing financial interests
%%
%\section{Competing financial interests}
%The authors declare no competing financial interests.
%\bibliography{bib}

\end{document}